\documentclass[aps,prl,reprint,groupedaddress]{revtex4-1}

\usepackage{graphicx}
\usepackage{dcolumn}
\usepackage{bm}

\begin{document}


\title{Parallelizing Wang-Landau algorithm in the field:\\
the micromagnetic ensemble}


\author{Borko D. Sto\v si\' c}
\email[]{borko@ufpe.br}
\affiliation{Departamento de Estat\' \i sica e Inform\' atica, 
Universidade Federal Rural de Pernambuco,\\
Rua Dom Manoel de Medeiros s/n, Dois Irm\~ aos,
52171-900 Recife-PE, Brazil}


\date{\today}

\begin{abstract}
It is shown in this work how the Wang-Landau algorithm
can be parallelized through the concept of the micromagnetic ensemble,
when the Hamiltonian contains both spin interaction and the external field terms, 
and thus energy-magnetization plane is
used for characterizing the density of states.
Within this framework random walk is performed on mutually independent
micromagnetic lines, and can thus be paralellized on a computer grid,
without need for shared memory among individual processes.
This approach pushes forward significantly the size of the
systems that may be addressed on current computer hardware
(from currently reported $42\times 42$ to at least $256\times 256$ in the 
case of two dimensional systems), 
and should turn out important for
diverse studies where field dependent behavior is essential.

\end{abstract}


\maketitle

Wang-Landau (WL) algorithm \cite{wang} represents the last (most successful) 
work in a series of attempts performed over the last couple of decades 
\cite{bhanot,swendsen,berg,lee,stinchcombe,oliveira,jwang}
to generalize the canonical importance sampling approach of Metropolis et al.
\cite{metropolis}. While importance sampling \cite{metropolis} 
traces out a path in the configurational space leading to equilibrium
configurations for the particular choice of external parameters
(such as temperature and field), all of these novel methods are
concerned with estimating the density of states (DOS), that is, the
number $g(E)$ of possible configurations available to the system at a
particular energy level $E$. The density of states curve is {\it
independent of} temperature, it depends on the topology of
the lattice alone, and it contains all the information necessary for
the complete solution of the problem at hand. 

The WL algorithm is accomplished through an iterative procedure, where a random walk is performed in the configurational space while simultaneously augmenting the density of states by a multiplicative factor $f>1$, and incrementing a histogram of visited configurations. The transition probability between states with energies $E$ and $E'$ is proportional to the ratio of the corresponding (previously accumulated) degeneracies $g(E)$ and $g(E')$, and thus by construction, more probable (higher entropy) 
energy levels develop higher DOS values $g(E)$, and transition probabilities level out so that a flat histogram is eventually attained. The multiplicative factor $f$ (typically starting out with a value of $\ln f=1$) is systematically reduced as $f\rightarrow f^{\frac{1}{4}}$ 
($\ln f\rightarrow \ln f-0.25$) 
and the histogram is reset every time that that it fulfills a (predefined) flatness criterion (common convention is when $80\%$ of visited states are not lower than $80\%$ of the histogram average), until a (predefined) lower bound of $f$ is reached 
(typically, $f_{min}=10^{-8}$). In short, WL is a highly intuitive, appealing, heuristic algorithm (with several somewhat arbitrary, but conventionally adopted parameters), which has been accepted by the scientific community as the state of the art, over the last decade or so. As at each step it (strongly) depends on the previous history of the ongoing simulation, WL may be regarded as a non-Marcovian Monte Carlo algorithm.

If the Hamiltonian contains a field term, the density of states $g(E,M)$ is 
defined on the energy-magnetization plane, and the simulation becomes far more 
demanding. In particular, while results for zero field have been reported \cite{wang} up to sizes $256\times 256$, memory requirements and the difficulty of convergence of the two dimensional random walk algorithm have up to date restricted studies of systems in a field to sizes $42\times 42$ \cite{wang2,wang3}. 
In comparison, exact results for DOS functions can be obtained for the zero field case using the method proposed by Beale \cite{beale} with algebraic manipulation software (such as Mathematica or Maple) up to size $64\times 64$ on current computer hardware, and exact DOS surfaces for the non-zero field case calculated using the transfer matrix method \cite{stosic} have been reported \cite{stosic2} up to size $12\times 12$.
In what follows, it is shown how system size limitation of the Wang Landau algorithm for systems in the field may  be significantly extended.

Without loss of generality, let us consider here the Ising model with  the nearest neighbor coupling $J$, in a uniform magnetic field $H$ (on an arbitrary lattice), with the Hamiltonian
\begin{equation}
{\cal H}=-J\sum_{<i,j>}S_i S_j-H\sum_i S_i \quad ,
\label{eq1}
\end{equation}
where $<>$ denotes summation over nearest neighbor pairs, and $S_i=\pm 1$ is the spin at site $i$.
The partition function of this system may be written as
\begin{equation}
Z=\sum_{k=0}^{N_B}\sum_{\ell=0}^{N}g_{k\ell} e^{-\beta E_{k\ell}}
,
\label{eq2} 
\end{equation}
where $N$ is the number of spins and $N_B$ the number of bonds, 
$E_{k\ell}=-J(N_b-2 k)-H(N-2\ell)$ is the energy of a configuration having
$k\in\{0,\dots N_b\}$ pairs of antiparallel spin pairs (number of ``unhappy" bonds), $\ell\in\{0,\dots N\}$ is the number of spins parallel to the field (``up" spins),
and $g_{k\ell}$ are the corresponding degeneracies.
Setting $H=0$ yields microcanonical degeneracies 
$g_k= \sum_{\ell}{g_{k\ell}}$, while setting $J=0$ 
leads to ``micromagnetic" degeneracies 
$g_{\ell}=\sum_{k}{g_{k\ell}}$, which correspond simply to the number of ways 
one can arrange $\ell$ ``up" spins in an N spin system, that is $g_{\ell}\equiv{N\choose \ell}$. 

The current approach is based on the observation that 
the Wang-Landau algorithm may be performed independently for individual micromagnetic lines $\ell$, as follows. The initial state $s_{\ell}(k)$ is prepared by flipping $\ell$ spins starting from an ordered state with all spins up, resulting in $k$ unhappy bonds, and lists of up-spin and down-spin indices are prepared. The initial density of states are set to $g_{\ell}(k)=1$, and the histogram values are set to zero $h_{\ell}(k)=0$. Lower index is used here to emphasize the fact that $\ell$ is fixed, and storage space corresponds 
{\it only to the number of bonds in the system} $N_B$
(rather than the product $N_B\times N$ of the number of bonds and the number of spins, as required by the
two dimensional WL random walk).
At each Monte Carlo step one spin is randomly chosen from the up-spin list and another from the down-spin list, and both spins are flipped to produce the proposal state $s_{\ell}(k')$ (therefore by construction proposed states preserve magnetization  - sampling is restricted within the current micromagnetic ensemble). 
The new (proposed) state is accepted with probability 
\begin{equation}
p\left(s_{\ell}(k)\rightarrow s_{\ell}(k')\right)=\min\left[ \frac{g_{\ell}(k)}{g_{\ell}(k')},1\right] \quad,
\label{eq3}
\end{equation}
where the density of states is modified by a multiplicative factor $f$ (initially $f=e$) to 
$g_{\ell}(k')\rightarrow g_{\ell}(k')*f$, histogram is incremented
$h_{\ell}(k')\rightarrow h_{\ell}(k')+1$, and the up-spin and down-spin lists are correspondingly updated. If the new state is rejected, the density of states of the current state and the corresponding histogram entries are updated to
$g_{\ell}(k)\rightarrow g_{\ell}(k)*f$ and $h_{\ell}(k)\rightarrow h_{\ell}(k)+1$,
respectively, while the up-spin and down-spin lists are not modified.
When the flatness criterion  is reached (e.g. of 80\% of histogram entries being greater than the histogram average), the histogram is reset, the density of states is normalized as 
\begin{equation}
g_{\ell}(k)\rightarrow g_{\ell}(k)\frac{{N\choose \ell}}{\sum_{j=0}^{N_B}{g_{\ell}(j)}}
\quad ,
\label{eq4}
\end{equation}
and the multiplicative factor is reduced to $f\rightarrow f^{\frac{1}4{}}$. 
The simulation stops when the multiplicative factor becomes less than some predefined value, here $f_{min}=1+N*10^{-9}$ is used, since the logarithm of the density of states scales as $N$ 
(the total number of configurations is $2^N$).

In fact, it was observed that at each new level of $f$ (after histogram reset together with DOS renormalization), the WL walker spends considerable time at the ends of the energy spectrum (the smallest entropy regions), and the histogram builds up rapidly in these regions, making it difficult to reach the flatness criterion. It was found that additional histogram resetting with density of states normalization at the same precision (unchanged value of $f$) is helpful in speeding up the algorithm, such that convergence with current precision is achieved in a matter of minutes for a single micromagnetic line for a $64\times 64$ system on a single core of a 2.4GHz Intel I7 processor, and in a matter of hours for a $256\times 256$ system

The current scheme is evidently ergodic, as any configuration with $\ell$ spins down
can be obtained from any other configuration that also has $\ell$ down spins, by flipping at most $\ell$ spin pairs (this worst case scenario corresponds to the situation when the Hamming distance between the two configurations assumes the maximum value of $\ell$). The WL random walk now becomes one dimensional, performed on the interaction energy levels corresponding to the current (chosen) micromagnetic ensemble. This fact, together with the reduced storage requirements, makes it possible to run much larger systems in comparison with the two dimensional walk.

\begin{figure}[h]
\centering
\begin{minipage}[b]{0.49\linewidth}
{\center{\hbox{\vbox{\includegraphics[width=1.75in]{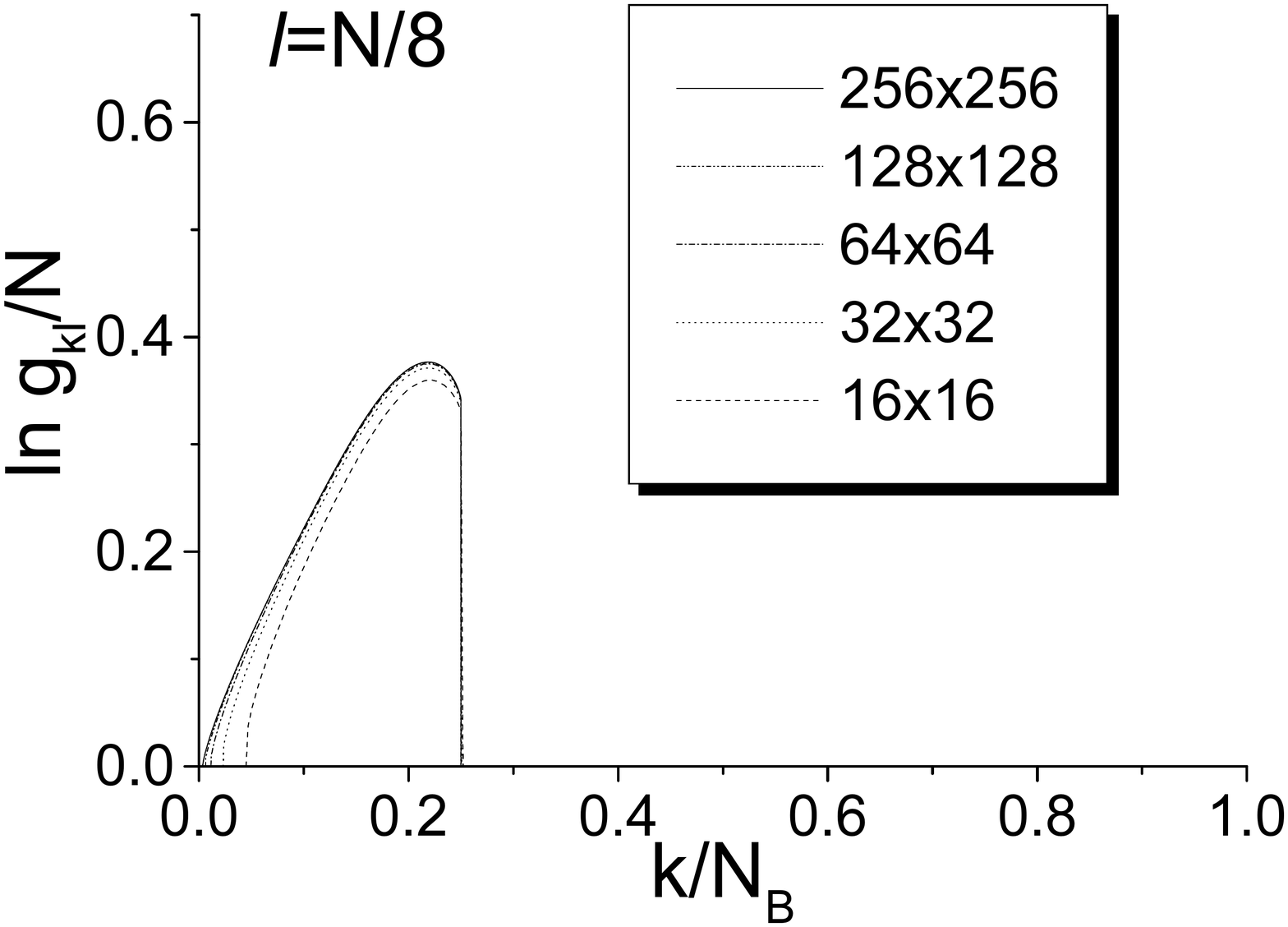}}}}}
\end{minipage}
\begin{minipage}[b]{0.49\linewidth}
{\center{\hbox{\vbox{\includegraphics[width=1.75in]{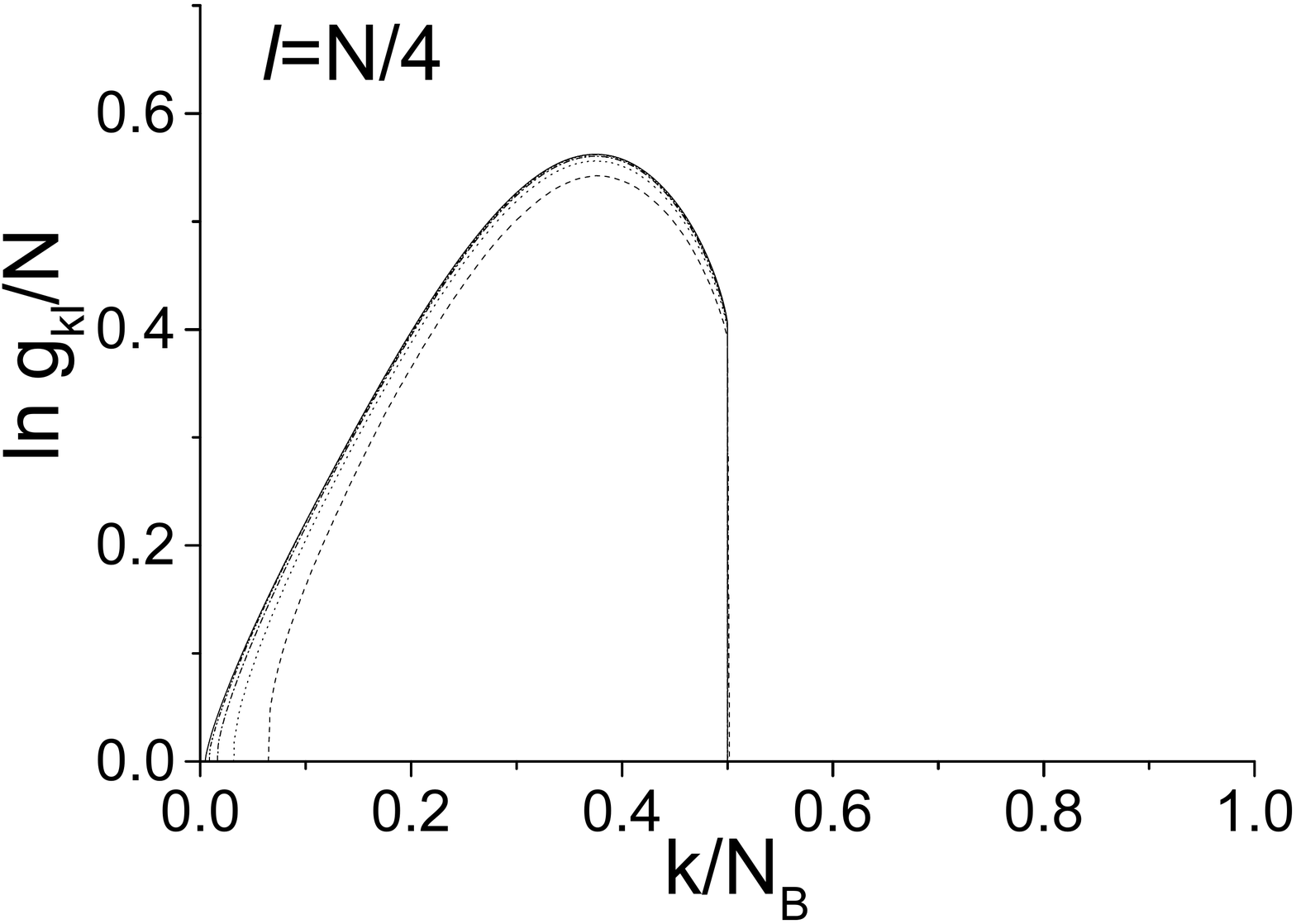}}}}}
\end{minipage}
\begin{minipage}[b]{0.49\linewidth}
{\center{\hbox{\vbox{\includegraphics[width=1.75in]{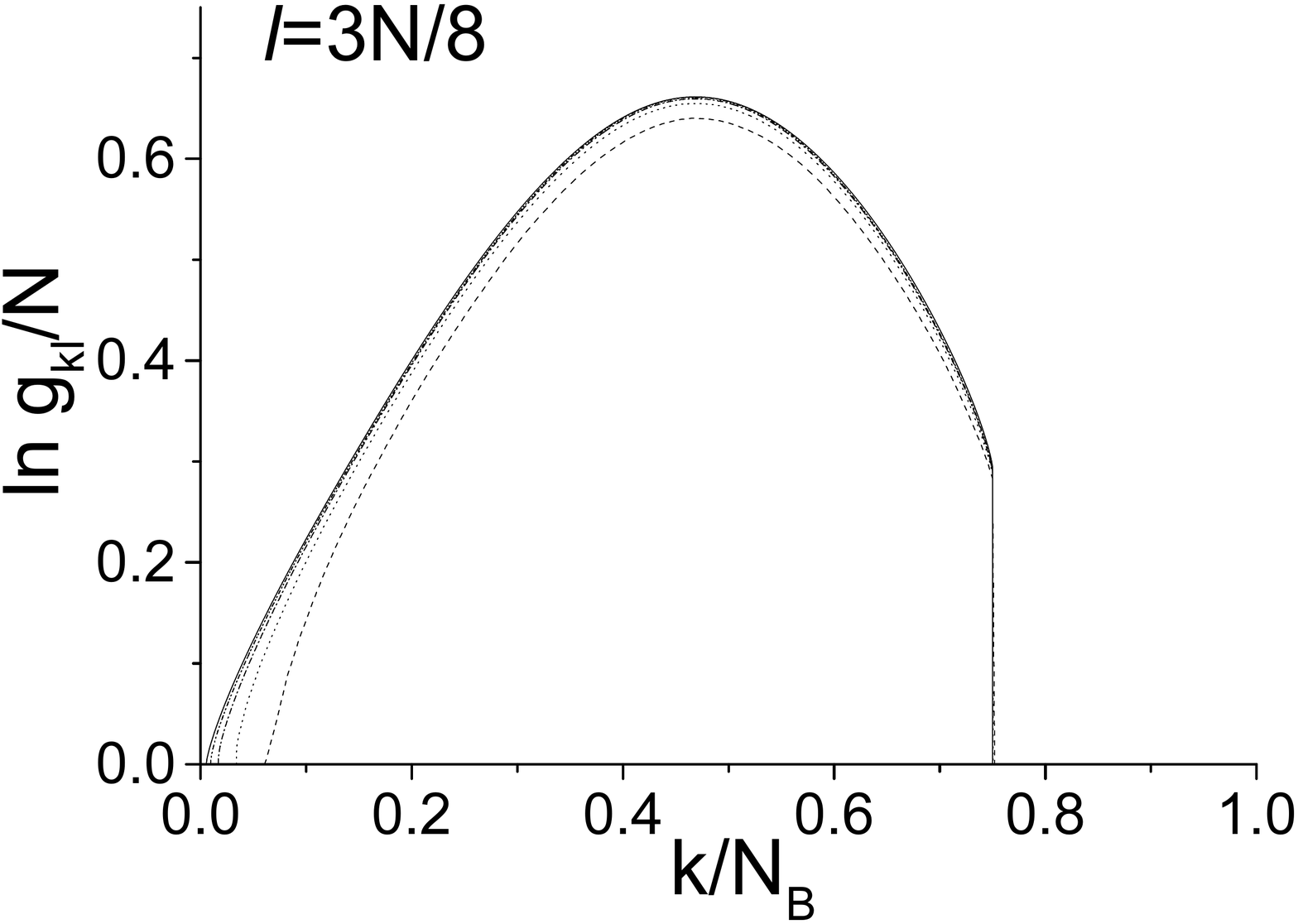}}}}}
\end{minipage}
\begin{minipage}[b]{0.49\linewidth}
{\center{\hbox{\vbox{\includegraphics[width=1.75in]{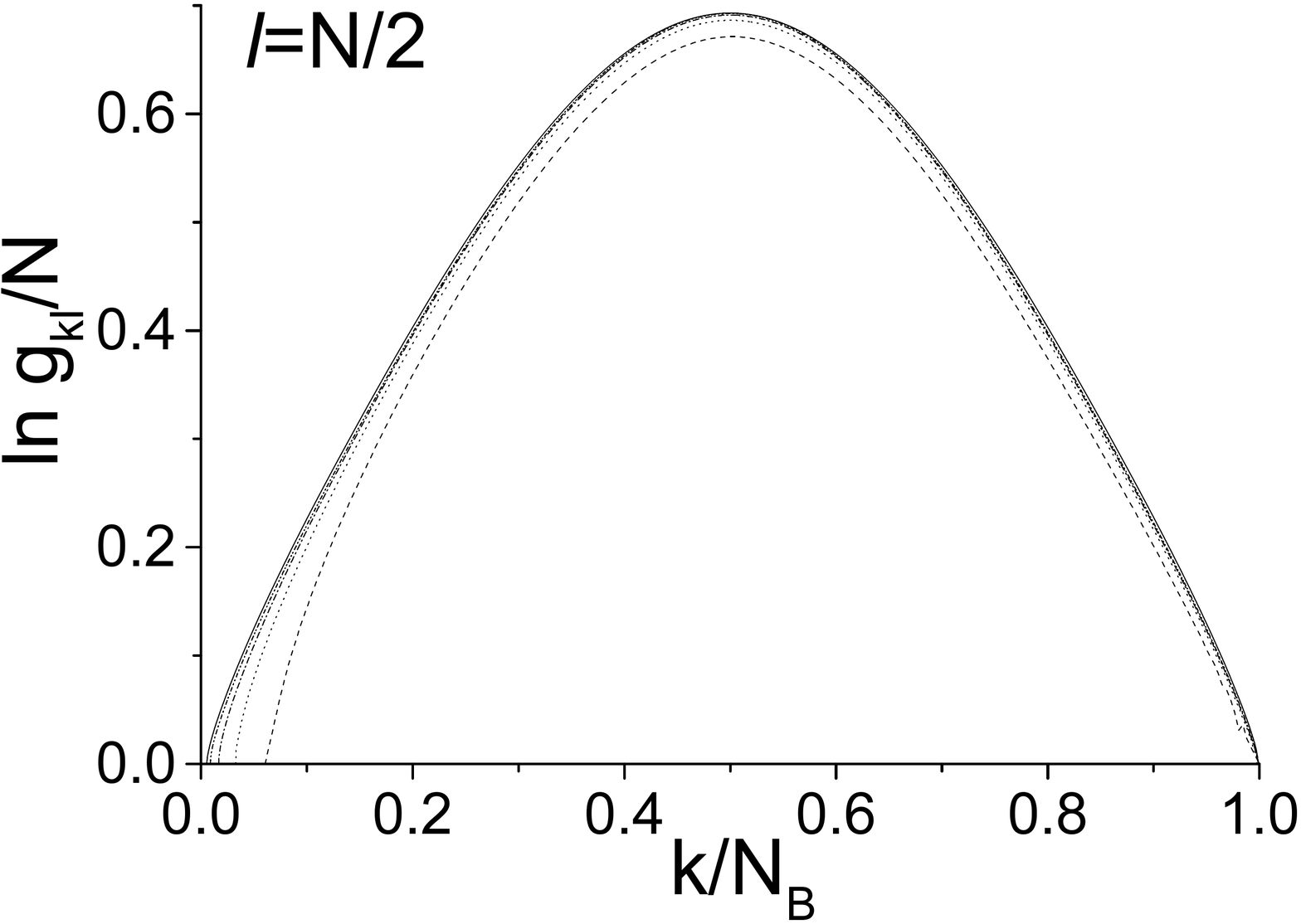}}}}}
\end{minipage}
\caption{Micromagnetic density of states 
$\ell=N/8$, $\ell=N/4$, $\ell=3 N/8$ and $\ell=N/2$,
for the $L\times L$ Ising model with periodic boundary conditions, 
for $L=16,32,64,128,256$.
}
\label{fig1}
\end{figure}

\begin{figure}
\centering
\begin{minipage}[b]{0.49\linewidth}
{\center{\hbox{\vbox{\includegraphics[width=1.75in]{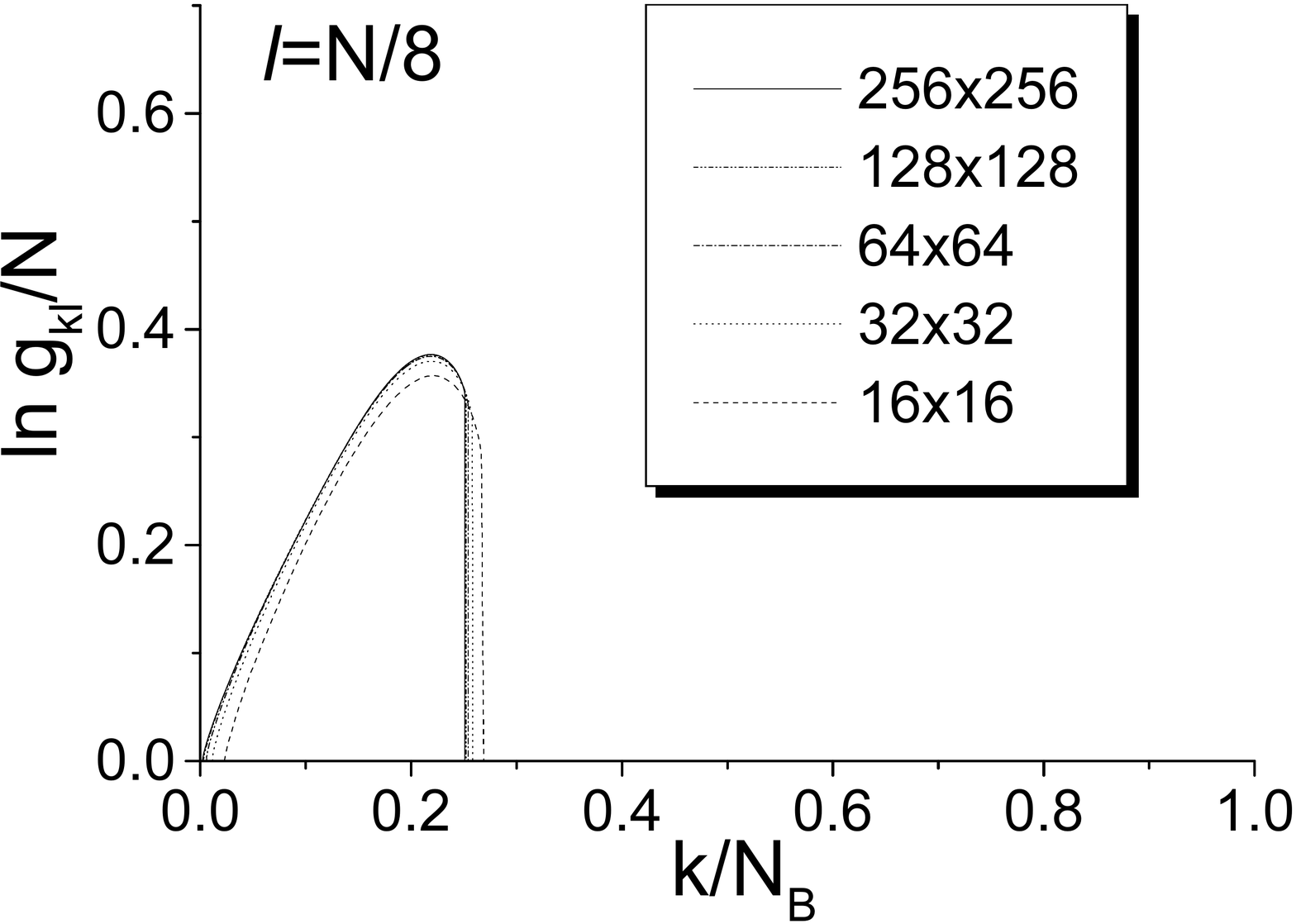}}}}}
\end{minipage}
\begin{minipage}[b]{0.49\linewidth}
{\center{\hbox{\vbox{\includegraphics[width=1.75in]{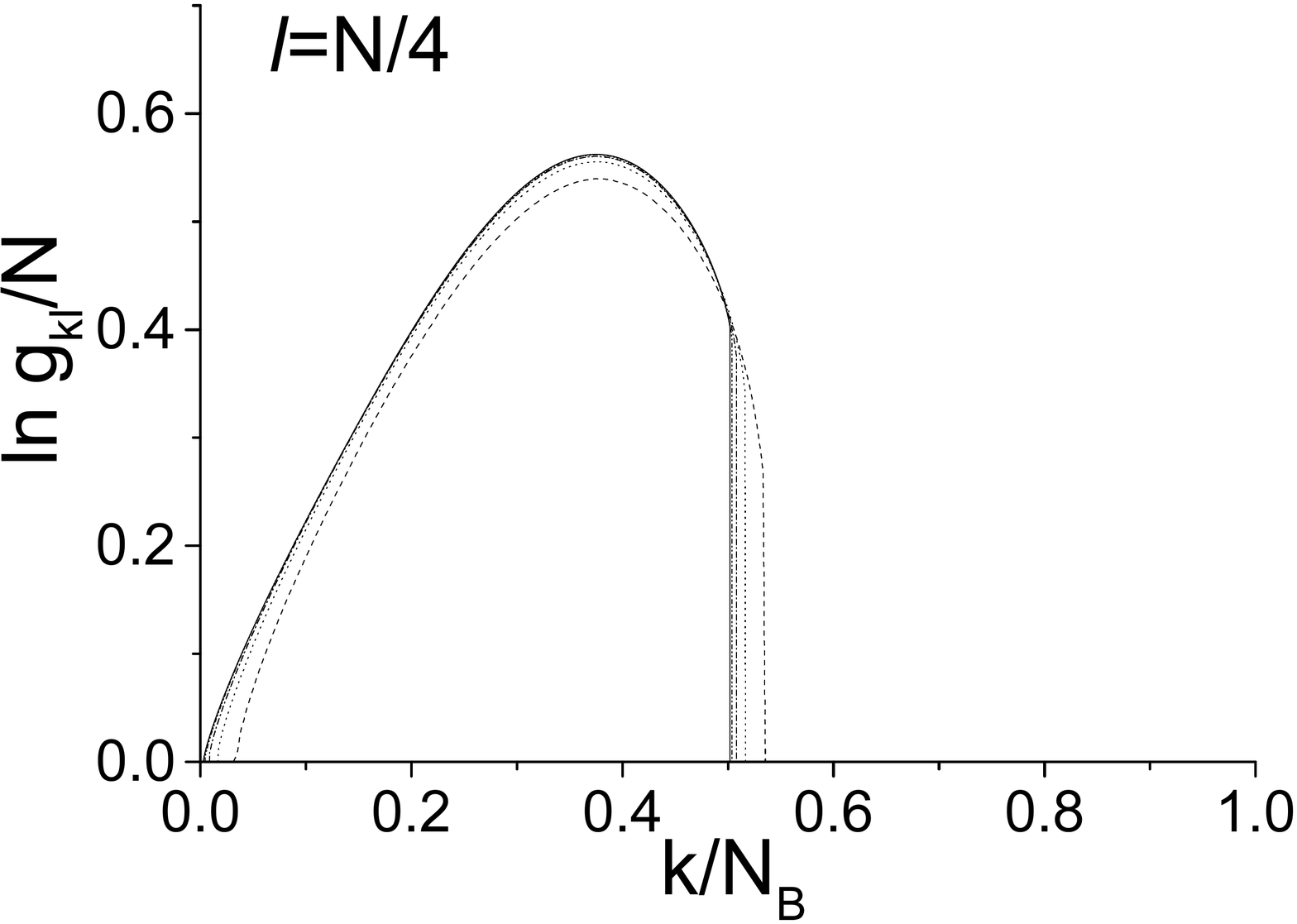}}}}}
\end{minipage}
\begin{minipage}[b]{0.49\linewidth}
{\center{\hbox{\vbox{\includegraphics[width=1.75in]{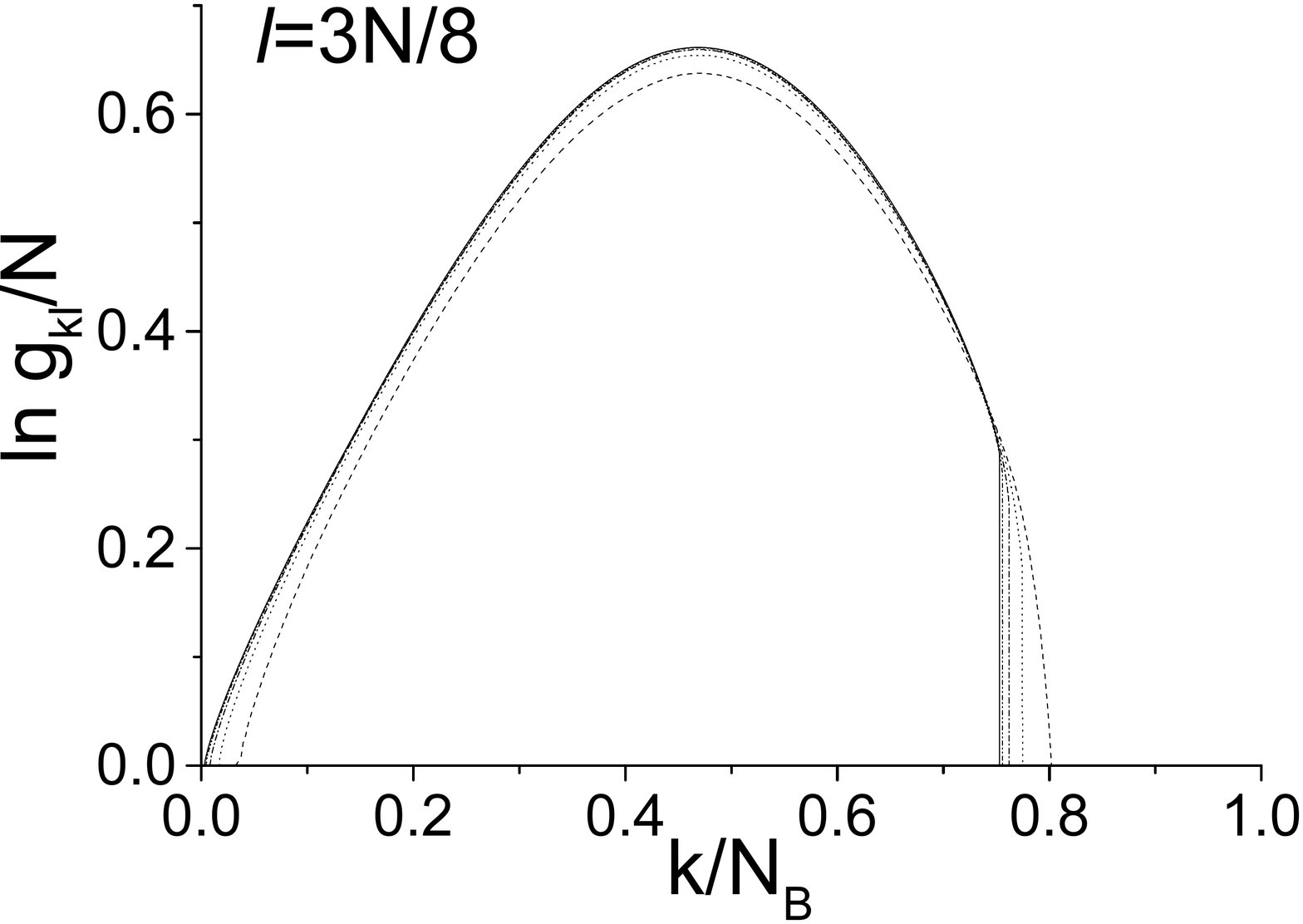}}}}}
\end{minipage}
\begin{minipage}[b]{0.49\linewidth}
{\center{\hbox{\vbox{\includegraphics[width=1.75in]{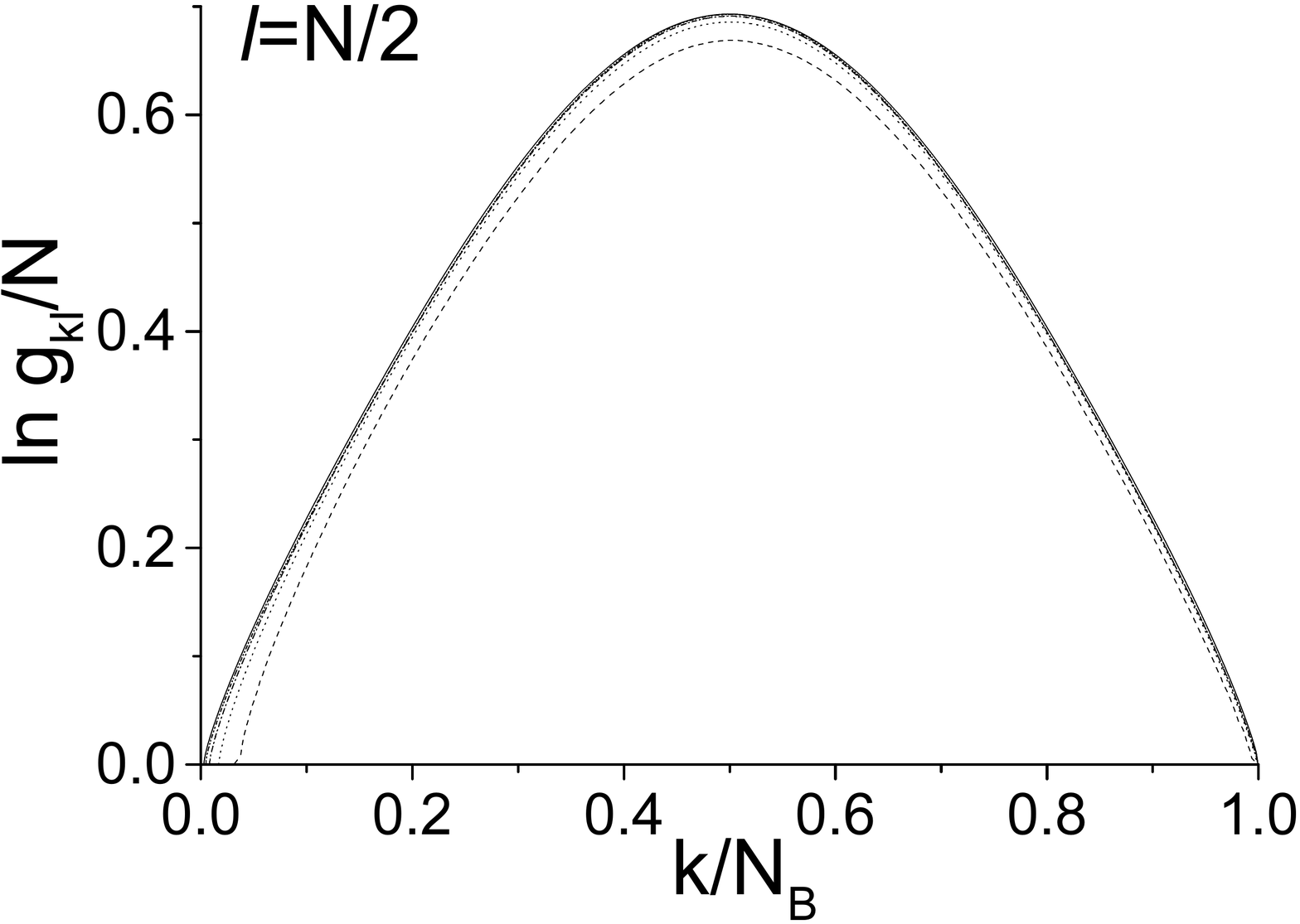}}}}}
\end{minipage}
\caption{Micromagnetic density of states 
$\ell=N/8$, $\ell=N/4$, $\ell=3 N/8$ and $\ell=N/2$,
for the $L\times L$ Ising model with open boundary conditions, 
for $L=16,32,64,128,256$.
}
\label{fig2}
\end{figure}

In Figs.~\ref{fig1} and \ref{fig2} results are presented for $L\times L$ Ising model systems for $L=16,32,64,128,256$ described by Hamiltonian (\ref{eq1}), for $\ell={N}/{8}, {N}/{4}, {3N}/{8}$ and ${N}/{2}$, which individually took up to 12 hours on a single core of an Intel I7 processor.

The difference between the way how the systems with periodic and open boundary conditions approach
the thermodynamic limit is evident on Figs.~\ref{fig1} and \ref{fig2}, 
and may be understood by considering the spin configurations that correspond to the low and high energy values.
The low energy configurations are accomplished by clustering of up and down spins, and for a periodic system with $N_B=2 L^2$ bonds two equal blocks of $N/2$ up and $N/2$ down spins are separated by an interface
consisting of $k=2 L$ frustrated bonds, such that the ground state corresponds to $k/N_B=1/L$ 
(e.g. $k/N_B=0.0625$ for $L=16$). 
For the systems with open boundaries with $N_B=2 L(L-1)$ bonds the two blocks are separated by an interface of $k=L$ frustrated bonds, so that the ground state corresponds to $k/N_B=1/2(L-1)$ 
(e.g. $k/N_B=0.0333$ for $L=16$). 

On the other side of the energy spectrum both periodic and open boundary systems with $N/2$ spins up and $N/2$ down spins have $k=N_B$ frustrated bonds in the Neel configurations, so that $k/N_B=1$.
For lower values of $\ell$ the high energy levels correspond to scattering of up spins, such that none are neighbors of each other, yielding $k=4 \ell$ frustrated bonds, and therefore the upper energy bound
is attained at $k/N_B=4 \ell/N_B$ ($k/N_B = N/2 N_B, N/N_B, 3 N/2 N_B$ for $\ell=N/8, N/4, 3 N/8$, respectively).
For periodic boundaries $N/N_B=1/2$ and the density of states curves on Fig.~\ref{fig1} end at $0.25, 0.5, 0.75$
for $\ell=N/8, N/4, 3 N/8$, while for open boundaries $N/N_B=L/2(L-1)$, and the limiting upper energy bound
levels are being gradually approached as $k/N_B = L/4 (L-1), L/2(L-1), 3L/4(L-1)$ for $\ell=N/8, N/4, 3 N/8$, respectively. 

\begin{figure}
\centering
\includegraphics[width=3.0in]{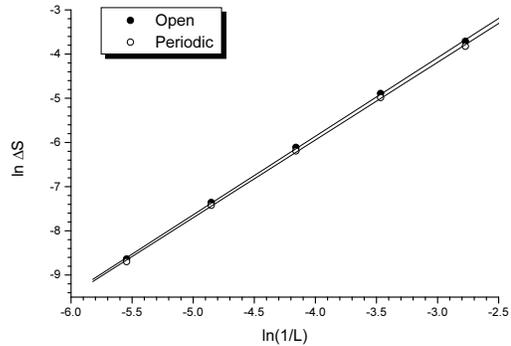}
\vspace{0.2cm}
\caption{
Difference of the scaled entropy value and the maximum value $\ln 2$
at the center of the energy magnetization plane, as a function of
inverse linear size, 
for the $L\times L$ Ising model with open and periodic boundary conditions, 
for $L=16,32,64,128,256$.
}
\label{fig3}
\end{figure}

Finite size scaling of the DOS curves evidently requires adjusting both axes (the dimensionless energy $U$ and the DOS function $S_{kl}\equiv \ln g_{k\ell}(L)$ scales), and does not appear to be a straightforward matter (studies in this direction are under way, and any conclusive results shall be reported elsewhere). On the other hand, general scaling behavior may be inferred by observing only the DOS function values $S_C(L)$ at the center of the energy magnetization plane ($k/N_B=0.5$, $\ell/N=0.5$), as a function of system size. In particular, the quantity $\Delta S(L) \equiv \ln 2-S_C(L)$ versus $\ln (1/L)$ (where $\ln 2$ is the limiting maximum entropy value) demonstrates linear behavior for both periodic and open boundary conditions, as shown in Fig.~\ref{fig3}.
Linear behavior of $\Delta S(L)$ versus $\ln (1/L)$ observed in Fig.~\ref{fig3} implies scaling form 
\begin{equation}
S_C(L)=\ln 2-\frac{\alpha}{L^{\beta}} ,
\label{eq5}
\end{equation}
where values obtained by regression are $\alpha=1.252$ and $\beta=1.778$ for open boundaries, while
for periodic boundary conditions parameter values $\alpha=1.088$ and $\beta=1.758$ are obtained.
For other points on the energy magnetization plane one may expect similar scaling behavior
\begin{equation}
S_{k\ell}(L)=S_{k\ell}(\infty)-\frac{\alpha_{k\ell}}{L^{\beta_{k\ell}}} ,
\label{eq6}
\end{equation}
where $S_{k\ell}(\infty)$ represents the corresponding entropy value in the thermodynamic limit.

It is seen that the micromagnetic DOS functions can be determined independently of each other using the current implementation of the WL algorithm, for very large systems. Nevertheless, determining the
complete set of micromagnetic lines that comprise the full DOS surface above the energy magnetization 
plane remains a formidable problem in terms of computer resources requirements, with a (rough) estimate of
200000 core-hours 
for the total of 32768 micromagnetic lines of a $256\times 256$ system (for models with up down symmetry, or double that number for asymmetric models), at the current precision of $f=1+N*10^{-9}$. On the other hand, as calculations for individual micromagnetic lines are fully independent of each other (no parameter region border adjustment, or relative normalization is needed), they may be performed on a geographically distributed computing grid, and the task may be regarded as rather demanding, but feasible.

\begin{figure}
\centering
\begin{minipage}[b]{0.49\linewidth}
{\center{\hbox{\vbox{\includegraphics[width=1.75in]{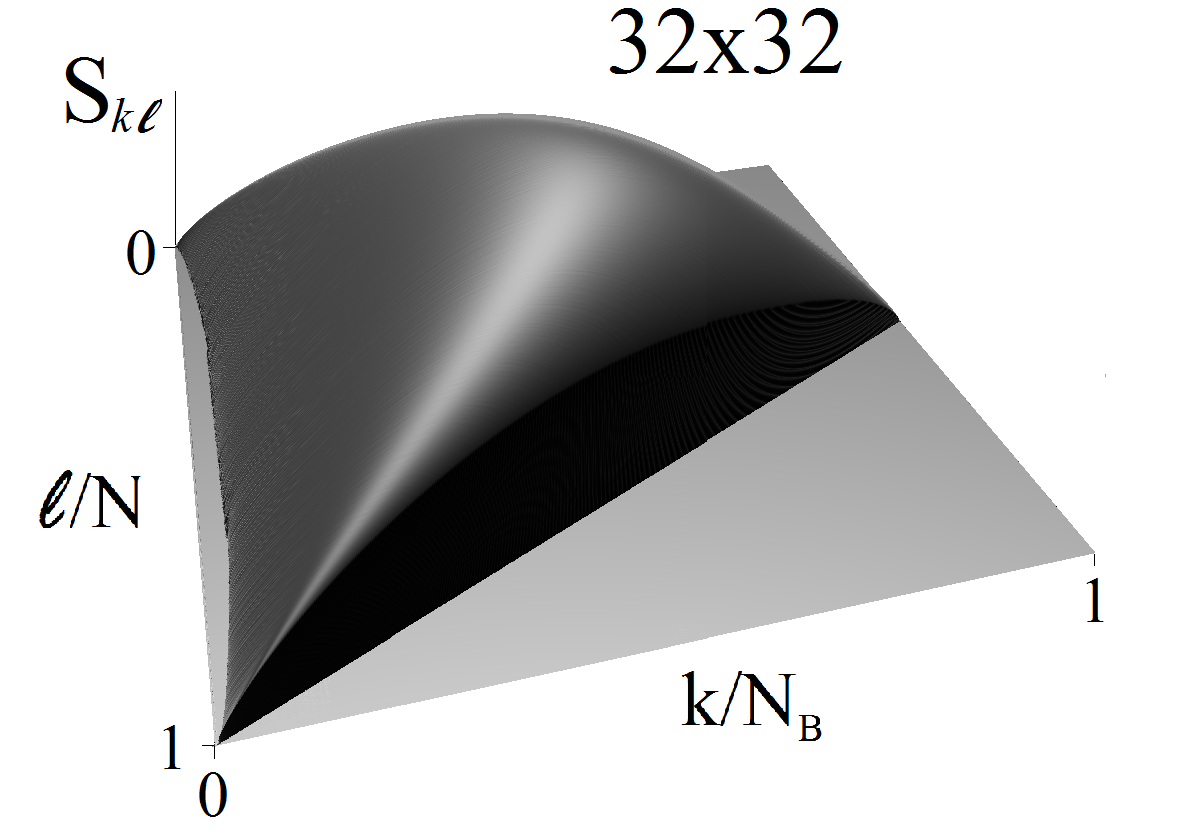}}}}}
\end{minipage}
\begin{minipage}[b]{0.49\linewidth}
{\center{\hbox{\vbox{\includegraphics[width=1.75in]{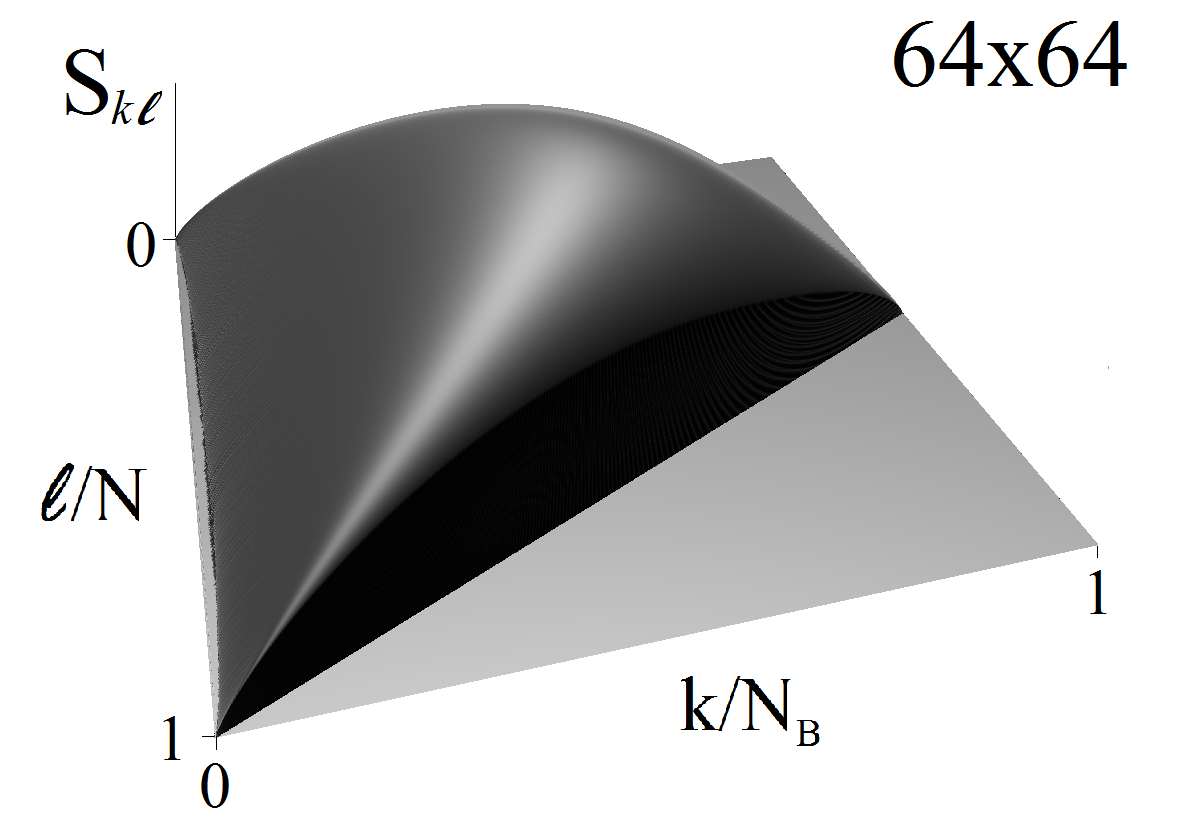}}}}}
\end{minipage}
\caption{Density of states for the $32\times 32$ and $64\times 64$ Ising model systems with periodic boundary conditions, calculated using the current implementation of the WL algorithm.
}
\label{fig4}
\end{figure}

The full DOS surfaces for $L\times L$ systems with periodic boundary conditions, for $L=32$ (obtained in a matter of hours on a single Intel I7 processor) and for $L=64$ (obtained in a matter of days) are shown in Fig.~\ref{fig4}.
In accordance with results displayed in Fig.~\ref{fig1}, it is seen on Fig.~\ref{fig4} that the DOS surfaces are rather similar, the largest differences being observed in the region of low energy ($k/N_B\sim 0$) and low absolute magnetization values ($\ell/N\sim 0.5$). The depression of the DOS surface in this region, surrounded by symmetric ridges that join smoothly with energy increase, is the signature of the second order transition.

\begin{figure}[h]
\centering
\begin{minipage}[b]{0.49\linewidth}
{\center{\hbox{\vbox{\includegraphics[width=1.75in]{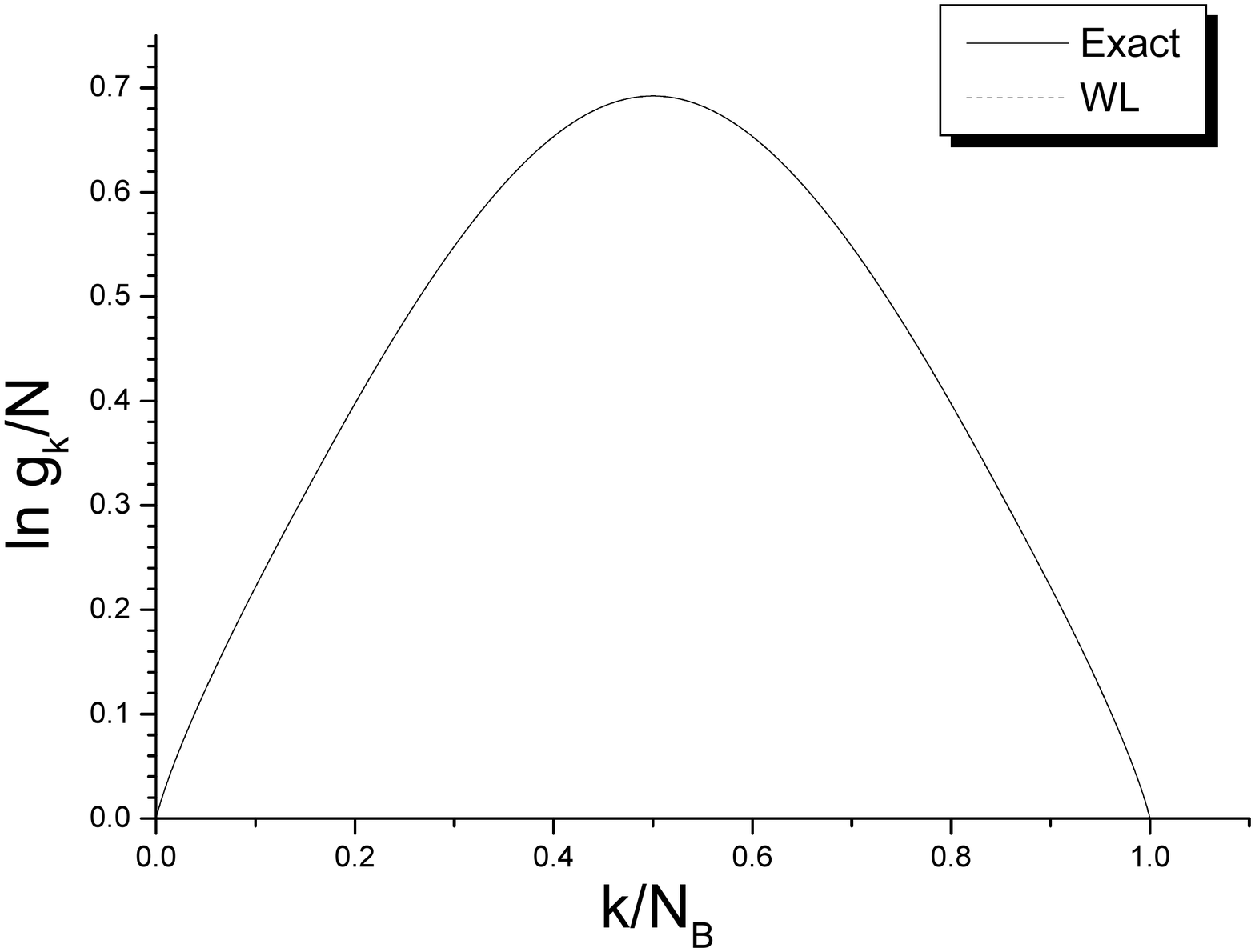}}}}}
\end{minipage}
\begin{minipage}[b]{0.49\linewidth}
{\center{\hbox{\vbox{\includegraphics[width=1.75in]{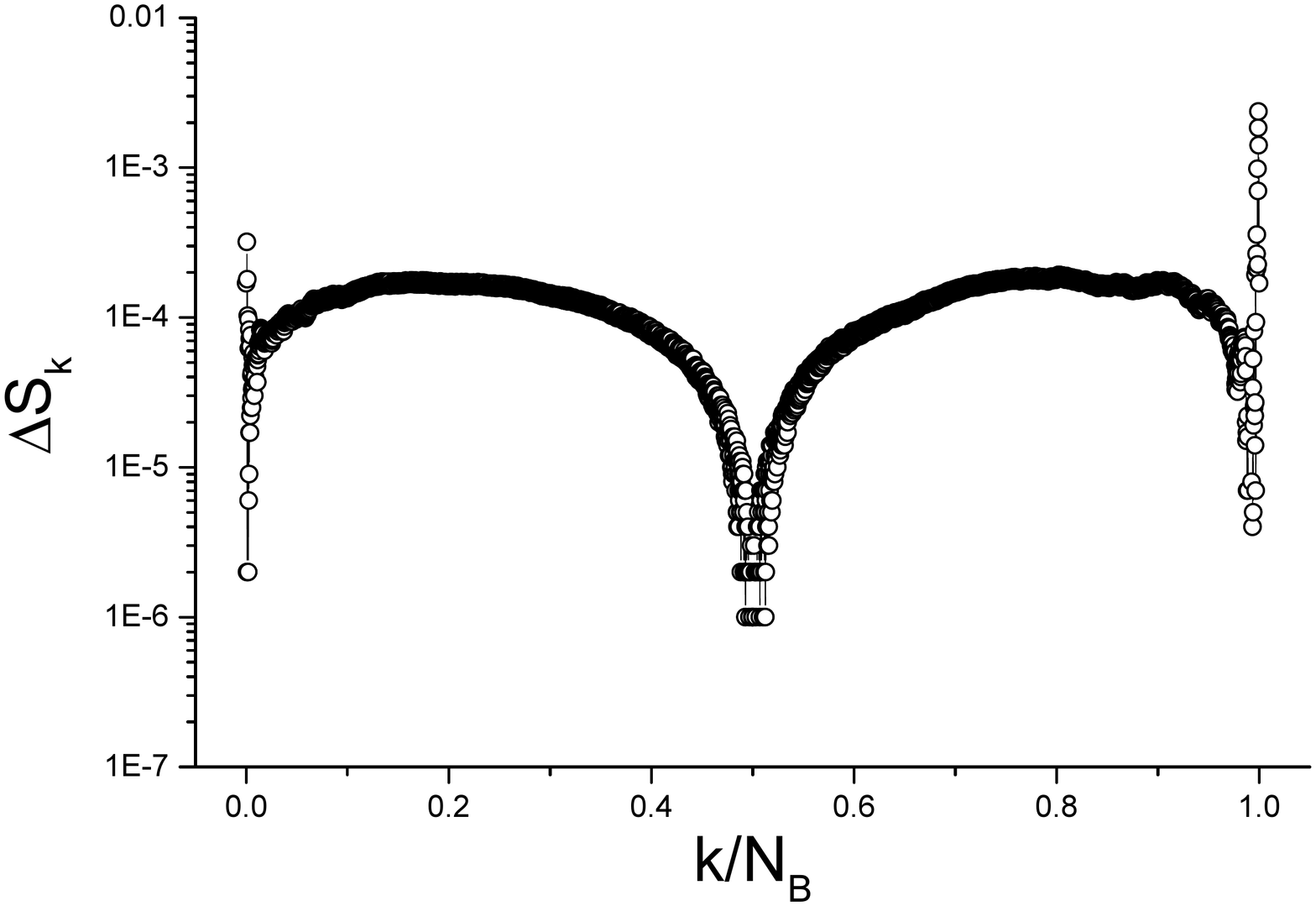}}}}}
\end{minipage}
\vspace{0.2cm}
\caption{
The exact density of states $S_k=\ln g_k/N_B$ for the $64\times 64$ Ising model with periodic boundary conditions, together with the results obtained by summing (over $\ell$) the micromagnetic curves obtained through
application of the Wang-Landau algorithm.
On the right side the difference between the exact and Wang-Landau algorithm results is shown.
}
\label{fig5}
\end{figure}

While the exact solution for the Ising model in the field is not known and there is no known exact result with which the DOS surfaces of Fig.~\ref{fig4} can be compared, one further test of validity can be made by summing the density of states over micromagnetic variable $\ell$, for each energy level, and than comparing the result with the exact density of states in zero field, obtained through the method proposed by Beale \cite{beale} using algebraic manipulation software (such as Mathematica or Maple).
Results of this comparison are displayed in Fig.~\ref{fig5}, where no visible difference is seen on the scale of the graph. 
The difference between the exact dos functions and the WL estimate shown on the right hand side of Fig.~\ref{fig5} is seen to be well below the value $\sqrt{\ln f} =\sqrt{N 10^{-9}}\sim 0.002$ \cite{zhou} (except at the very ends of the energy spectrum range), which may be attributed \cite{zhou} to multiple simulations performed for distinct micromagnetic ensembles (different values of $\ell$), that were used here to compose the microcanonical entropy curve.

In summary, in this work it is shown how the Wang Landau algorithm may be optimized for systems with a field dependent Hamiltonian,
by running independent micromagnetic ensemble runs. Updates are performed by conserving magnetization (number of spins parallel to the field), and this reduction of the dimension of the parameter space brings about considerable advantages.
First, memory requirements are reduced from $N_B\times N$ (where $N_B$ is the number of bonds, and $N$ number of spins in the system) to $N_B$. Second, the micromagnetic ensemble runs are fully independent of each other and may be performed in parallel on a geographically distributed grid. And finally, convergence of the algorithm is observed on much larger systems as compared with a two dimensional WL random walk (pushing the limit from $42\times 42$ reported field results \cite{wang2,wang3} to at least $256\times 256$). This approach is independent of the details of geometry and interactions
(although examples presented here were performed on a nearest neighbor Ising ferromagnet), as long as the uniform field term is present in the Hamiltonian.

\begin{acknowledgments}
This work was supported by CNPq (Brazilian Agency).
\end{acknowledgments}


\end{document}